\documentclass[pre,showpacs,aps,amsmath]{revtex4}
\usepackage{graphicx}
\usepackage{epsfig}

\def\text#1{\mbox{#1}}

\def\be#1{\begin{equation}\label{#1}}
\def\ee{\end{equation}}

\usepackage{graphicx}
\usepackage{dcolumn}
\usepackage{bm}

\begin{document}

\title{
  Dualization of non-Abelian $B\wedge\varphi$ model
}

\author{C. A. S. Almeida}
 \email{carlos@fisica.ufc.br}
\affiliation{
  Departamento de F\'{\i}sica, Universidade Federal do Cear\'{a},\\
  Caixa Postal 6030, 60455-760 Fortaleza, Cear\'{a}, Brazil}

\author{R. R. Landim}
\affiliation{
  Departamento de F\'{\i}sica, Universidade Federal do Cear\'{a},\\
  Caixa Postal 6030, 60455-760 Fortaleza, Cear\'{a}, Brazil}

\author{D. M. Medeiros}
 \email{dedit@fisica.ufc.br}
\affiliation{
  Departamento de F\'{\i}sica, Universidade Federal do Cear\'{a},\\
  Caixa Postal 6030, 60455-760 Fortaleza, Cear\'{a}, Brazil}
\affiliation{ Universidade Estadual do Cear\'{a} - Departamento de
F\'{\i}sica e Qu\'{\i}mica \\Av. Paranjana, 1700, 60740-000
Fortaleza-Ce, Brazil}

\begin{abstract}
In this work we show a dualization process of a non-Abelian model
with an antisymmetric tensor gauge field in a three-dimensional
space-time. We have constructed a non-Abelian gauge invariant
St\"{u}ckelberg-like master action, and a duality between a
non-Abelian topologically massive $B\wedge\varphi$ model and a
non-Abelian massive scalar action, which leads us to a
Klein-Gordon-type action when we consider a particular case.
\end{abstract}



\maketitle

\section{Introduction}
Interest has recently increased in dualization of non-Abelian
theories. Indeed, Smailagic and Spallucci, in the framework of
gauge models, have investigated a dualization of a non-Abelian
$B\wedge F$ model in $D=4$, in order to obtain a St\"{u}ckelberg
massive gauge invariant theory \cite
{smailagic1,smailagic2,smailagic3}. It is worth to mentioning that
the dualization of topological models \cite{karl} as well as the
interest for St\"{u}ckelberg-like gauge invariant models is partly
due to its relevance as alternatives to the Higgs mechanism for
gauge fields mass generation \cite {jackiw,allen,khoudeir}.

The method used was first introduced by Buscher \cite{buscher}.
The procedure consists of gauging a symmetry of the original
action by introducing non-propagating gauge fields and
constraining the respective field strength to vanish by means of a
Lagrange multiplier. After integrating over the Lagrange
multiplier and fixing the gauge, we recover the original action.
On the other hand, by integrating by parts the Lagrange multiplier
term and then integrating out the gauge fields, we obtain the dual
action.

In the framework of non-linear sigma models with non-Abelian
isometries, the developments started with the work of de la Ossa
and Quevedo \cite{quevedo}, after a dual theory of an arbitrary
sigma model with an Abelian isometry has been constructed
\cite{verlinde}. More recently, Mohammedi shed new lights about
this issue \cite{mohammedi1,mohammedi2}.

Recently we have discussed an Abelian three-dimensional action
with a topological term involving a two-form gauge field $B$ and a
scalar field, called for short $B\wedge\varphi$ term, in the
framework of topological mass generation. Also we have showed that
this action is related by Buscher's duality transformation to a
massive gauge-invariant St\"{u}ckelberg-type theory \cite{dedit1}.

This is the only one model involving a Kalb-Ramond field defined
in (2+1) dimensions. Besides the importance of dualities for
string theories, the issue of broken symmetries restored at the
quantum level could be enlightened by duality procedures. Indeed,
a massive gauge invariant theory and a dual theory with the gauge
symmetry broken may bring us interesting information at the
quantum level.

\section{The Abelian master action}

In this letter we will give the dualization of a non-Abelian version of the $%
B\wedge\varphi$ topological model. But it is instructive to
analyse first the procedure for the Abelian theory, where the
master action in three dimensions \cite{dedit1} is given by

\begin{equation}
S =\int_{M_{3}}\left\{ \frac{1}{2}H\wedge ^{\ast
}H+\frac{1}{2}(d\phi -U)\wedge ^{\ast }(d\phi -U)+mB\wedge (d\phi
-U)+ m\Gamma \wedge dU\right\} . \label{det01}
\end{equation}

In the expression above, $U$ is the St\"{u}ckelberg auxiliary
vectorial field, $\Gamma$ is the Lagrange multiplier field, and
$H\equiv dB$ is a three form field-strength of the so-called
Kalb-Ramond field $B$. The action (\ref{det01}) above has
invariance under transformations

\begin{equation}
\delta B=d\Omega ,\quad \quad \delta U=d\lambda ,\quad \quad
\delta \phi =\lambda ,\quad \quad \delta \Gamma =d\alpha.
\label{det02}
\end{equation}

Variation of the action (\ref{det01}) above with respect to
$\Gamma$ gives the following field equation of motion:

\begin{equation}
dU=\alpha . \label{det03}
\end{equation}
Therefore, from Poincar\'{e}'s lemma, the auxiliary field is an
exterior derivative of a zero form, namely,

\begin{equation}
U=d\phi ^{\prime }. \label{det04}
\end{equation}
Putting the results (\ref{det03}) and (\ref{det04}) in the master
action (\ref{det01}) and defining a new scalar field $\varphi
=\phi -\phi ^{\prime}$ , we have a $B\wedge \varphi $ model
described by the action

\begin{equation}
S_{B\varphi }=\int_{M_{3}}\left\{ \frac{1}{2}H\wedge ^{\ast }H+\frac{1}{2}%
d\varphi \wedge ^{\ast }d\varphi +mB\wedge d\varphi \right\} .
\label{det05}
\end{equation}
Since $\phi ^{\prime }$ arises from the St\"{u}ckelberg auxiliary
field as
given in the equation (\ref{det04}), a new definition of the scalar field $%
\phi $ causes no change in this physical problem. Therefore, the
action (\ref {det05}) is gauge invariant under the following
transformations
\begin{equation}
\delta B=d\Omega ,\quad \quad \delta \varphi =0. \label{det06}
\end{equation}
Now, the variation in (\ref{det01}) with respect to the
Kalb-Ramond field $B$ implies that

\begin{equation}
d^{\ast }H-m\left( d\phi -U\right) =0. \label{det07}
\end{equation}
However, when we consider both the equation (\ref{det04}) and the
definition of the scalar field $\varphi $ in the equation
(\ref{det07}) we obtain the equation

\begin{equation}
d\left( ^{\ast }H-m\varphi \right) =0, \label{det08}
\end{equation}
whose general solution is given by

\begin{equation}
^{\ast }H-m\varphi =\Phi . \label{det09}
\end{equation}
Inserting the former solution (\ref{det09}) in the master action
(\ref{det01}) we find a massive model described by the action

\begin{equation}
S_{\Phi }=\int_{M_{3}}\left\{ \frac{1}{2}d\varphi \wedge ^{\ast }d\varphi -%
\frac{m}{2}\varphi \wedge ^{\ast }\left( m\varphi +\Phi \right) +\frac{1}{2}%
\Phi \wedge ^{\ast }\left( m\varphi +\Phi \right) \right\} .
\label{det10}
\end{equation}

It is worth to mentioning that the scalar field $\Phi$ in the
action (\ref{det10}) has no propagation. Besides, if we consider
the particular case where $^{\ast }H-m\varphi $ is constant and
zero, or in other words, if we consider the equations of motion,
we obtain the Klein-Gordon massive model, namely

\begin{equation}
S_{KG}=\int_{M_{3}}\left\{ \frac{1}{2}d\varphi \wedge ^{\ast }d\varphi -%
\frac{m^{2}}{2}\varphi \wedge ^{\ast }\varphi \right\} ,
\label{det1011}
\end{equation}
where we have used
\begin{equation}
^{\ast }H=m\varphi , \label{det1012}
\end{equation}
which is the particular case mentioned above.

Therefore, from the master action (\ref{det01}) we have shown that
the $B\wedge \varphi $ model (\ref{det05}) and the  action for a
massive scalar field (\ref{det10}) are dual to each other. Also we
have shown that the action (\ref{det10}) leads to a free and
massive Klein-Gordon action, when we consider the particular case
where $^{\ast }H-m\varphi $ is constant and zero.

\section{The Non-Abelian master action}

Next, we shall see that there is a simple extension of the above
procedure in the case of a non-Abelian internal symmetry. It is
interesting to remark that the only possibility to construct a
non-Abelian version of the master action (\ref{det01}) is via an
introduction of an auxiliary vector field, as we have proved in
Ref. \cite{dedit2}, using the method of consistent deformations.
Also, it is important to point out that the introduction of a one
form gauge connection A is required to go further in the
non-Abelian generalization of our model, although our original
Abelian action (\ref {det01}) does not contain this field. Note
that, as pointed out by Thierry-Mieg and Neeman \cite{thierry} for
the non-Abelian case, the field strength for B is

\begin{equation}
H=dB+\left[ A,B\right] \equiv DB  \label{det11a}
\end{equation}

Following Ref. \cite{thierry}, we can define a new $\mathcal{H}$
given by
\begin{equation}
\mathcal{H}=DB+\left[ F,\Lambda \right] \label{det12}
\end{equation}
where $\Lambda $ is a one form auxiliary field and $F=dA+A\wedge
A$.

The obstruction to the non-Abelian generalization lies only on the
kinetic term for the antisymmetric field, but the topological term
must be conveniently redefined. Therefore our non-Abelian master
action can be written as \cite{dedit3}

\begin{equation}
S =\int_{M_{3}}\left\{ \frac{1}{2}\mathcal{H}\wedge ^{\ast
}\mathcal{H}+ \frac{1}{2}(D\phi -U)\wedge ^{\ast }(D\phi
-U)+mB\wedge (D\phi -U)+m\Gamma \wedge DU\right\} , \label{det13}
\end{equation}

The action (\ref{det13}) is invariant under the following
transformations
\begin{equation}
\delta A=-D\theta ,\quad \delta \phi =\left[ \theta ,\phi \right]
,\quad
\delta B=D\Omega +\left[ \theta ,B\right] ,\quad \delta \Lambda =\Omega +%
\left[ \theta ,\Lambda \right] , \label{det14}
\end{equation}
where $\theta $ and $\Omega $ are zero and one form
transformations parameters, respectively. The equation of motion
with respect to the Lagrange multiplier field provides an
analogous constraint to the obtained in the Abelian case (eq.
(\ref{det03})). This is possible only when a ''flat connection''
is imposed for the gauge vectorial field, that is $F(A)=0$.
Therewith, the variation of the model (\ref{det13}) regarding
$\Gamma $ supply us with
\begin{equation}
DU=0. \label{det15}
\end{equation}
Thus we have now
\begin{equation}
U=D\phi ^{\prime }. \label{det16}
\end{equation}
Again the field $\phi ^{\prime }$ is a zero form as in the
previous case (eq.(\ref {det04})). Going back to the equation
(\ref{det13}) with the results (\ref {det15}) and (\ref{det16})
and using the same former definition for a new scalar field
$\varphi =\phi -\phi ^{\prime }$, that causes no change in the
physical model, we obtain a non-Abelian $B\wedge \varphi $ model,
namely,
\begin{equation}
S_{B\varphi }^{^{\prime }}=\int_{M_{3}}Tr\left\{ \frac{1}{2}\mathcal{H}%
\wedge ^{\ast }\mathcal{H}+\frac{1}{2}D\varphi \wedge ^{\ast }D\varphi -m%
\mathcal{H}\wedge \varphi \right\} . \label{det17}
\end{equation}

Continuing with the same procedure of the Abelian case, we vary the action (%
\ref{det13}) with respect to the two form $B$ and obtain
\begin{equation}
D\left( ^{\ast }\mathcal{H}-m\varphi \right) =0, \label{det18}
\end{equation}
which leads us to
\begin{equation}
^{\ast }\mathcal{H}-m\varphi =\xi, \label{det19}
\end{equation}
Again, inserting the solution (\ref{det19}) in the master action
(\ref {det13}) we have a non-Abelian massive action, namely,
\begin{equation}
S_{\xi}^{^{\prime }}=\int_{M_{3}}Tr\left\{ \frac{1}{2}D\varphi
\wedge ^{\ast }D\varphi -\frac{m}{2}\varphi \wedge ^{\ast }\left( m\varphi +%
\xi\right) +\frac{1}{2}\xi\wedge ^{\ast }\left( m\varphi
+\xi\right) \right\} , \label{det20}
\end{equation}
which is similar to the Abelian case.

Note that, analogous to the Abelian case, in the model
(\ref{det20}) the scalar field $\xi$ has not propagation and using
the particular case where $^{\ast }\mathcal{H}-m \varphi $ is
constant and zero we obtain a kind of non-Abelian Klein-Gordon
massive model, namely
\begin{equation}
S_{KG}^{^{\prime }}=\int_{M_{3}}Tr\left\{ \frac{1}{2}D\varphi
\wedge ^{\ast }D\varphi -\frac{m^{2}}{2}\varphi \wedge ^{\ast
}\varphi \right\} . \label{det2011}
\end{equation}

\section{Conclusions}

In summary, we have shown in this letter a dualization process of
a non-Abelian model in a three-dimensional space-time. We have
constructed a non-Abelian gauge invariant master action, which
generates a non-Abelian St\"{u}ckelberg-like $B\wedge \varphi $
model and a non-Abelian massive action. For that, we have used the
well established Buscher\'{}s dualization method.

Kalb-Ramond fields arise naturally in string theory coupled to the
area element of the two-dimensional worldsheet \cite{kalb}. It is
worthwhile to mentioning that duality between models involving
Kalb-Ramond fields and scalar ones is rather rare in the
literature.

\vspace{0.3in} \centerline{\bf ACKNOWLEDGMENTS} DMM would like to
thank Dr. Manasses Claudino Fonteles. This work was supported in
part by Conselho Nacional de Desenvolvimento Cient\'{\i }fico e
Tecnol\'{o}gico-CNPq.



\begin{references}

\bibitem{smailagic1}  A. Smailagic, E. Spallucci, \emph{Phys. Lett.} \textbf{%
B 489 }(2000) 435.

\bibitem{smailagic2}  A. Smailagic, E. Spallucci, \emph{Phys. Rev.} \textbf{%
D61} (2000) 067701.

\bibitem{smailagic3}  A. Smailagic, E. Spallucci, \emph{Phys. Lett.} \textbf{%
B 471 }(1999) 133.

\bibitem{karl}  A. Karlhede, U. Lindstr\"{o}m, M. Ro\v{c}ek, P. Van
Nieuwenhuizen, \emph{Phys. Lett.} \textbf{B 186 }(1987) 96.

\bibitem{jackiw}  S. Deser, R. Jackiw, and S. Templeton, \emph{Ann. Phys.}
\textbf{140} (1982) 372.

\bibitem{allen}  T. J. Allen , M. J. Bowick and A. Lahiri, \emph{Mod. Phys.
Lett.}\textbf{A6} (1991) 559; A. Lahiri, Generating vector boson
masses,
hep-th/9301060; M. LeBlanc, R. MacKenzie, P.K. Panigraghi, and R. Ray, \emph{%
Int. J. Mod. Phys.} \textbf{A9} (1994) 4717.

\bibitem{khoudeir}  A. Khoudeir, \emph{Phys. Rev.} \textbf{D 59} (1999)
027702.

\bibitem{buscher}  T. Buscher, \emph{Phys. Lett.} \textbf{B 159 }(1985) 127;
\emph{Phys. Lett.} \textbf{B 194 }(1987) 59; \emph{Phys. Lett.}
\textbf{B 201 }(1988) 466.

\bibitem{quevedo}  X. C. de la Ossa, F. Quevedo, \emph{Nucl. Phys.} \textbf{%
B 403 }(1993) 377.

\bibitem{verlinde}  M. Ro\v{c}ek, E. Verlinde, \emph{Nucl. Phys.} \textbf{B
373 }(1992) 630.

\bibitem{mohammedi1}  N. Mohammedi, \emph{Phys. Lett.} \textbf{B 375 }(1996)
149.

\bibitem{mohammedi2}  A. Bossard, N. Mohammedi \emph{Nucl. Phys.} \textbf{B
595 }(2001) 93.

\bibitem{dedit1}  D. M. Medeiros, R. R. Landim, C. A. S. Almeida, \emph{%
Europhys. Lett.~}\textbf{48} (1999) 610.

\bibitem{dedit2}  D. M. Medeiros, R. R. Landim, C. A. S. Almeida, \emph{%
Phys. Lett.~}\textbf{B 502} (2001) 300.

\bibitem{thierry}  J. Thierry-Mieg and Y. Neeman, \emph{Proc. Natl. Acad.
Sci. U.S.A.~} \textbf{79} (1982) 7068.

\bibitem{dedit3}  D. M. Medeiros, R. R. Landim, C. A. S. Almeida, \emph{%
Phys. Rev.} \textbf{D 63} (2001) 127702.

\bibitem{kalb}  M. Kalb and P. Ramond, \emph{Phys. Rev.} \textbf{D 9} (1974)
2273.

\end{references}
\end{document}